# Comments on the paper: 'Growth structural, spectral, optical and mechanical studies of gamma bis glycinium oxalate (GBGOx) new NLO single crystal by SEST method'

Bikshandarkoil R. Srinivasan, Kiran T. Dhavskar
Department of Chemistry, Goa University, Goa 403206, INDIA
Email: srini@unigoa.ac.in Telephone: 0091-(0)832-6519316; Fax: 0091-(0)832-2451184

**Abstract**

The authors of the title paper (Optik, 125 (2014) 1825-1828) claim to have synthesized a new nonlinear optical (NLO) gamma bis glycinium oxalate (GBGOx) crystal by slow evaporation solution technique. In this communication, many points of criticism, concerning the characterization of this so called GBGOx NLO crystal are highlighted to prove that the title paper is completely erroneous.

**Introduction**

Glycine an achiral amino acid, represented by the zwitter ionic formula $^+NH_3$-$CH_2$-$COO^-$ exists in three polymorphic modifications namely α- or β- or γ-glycine [1-3]. In accordance with its achiral nature, a majority of the known structurally characterized compounds of glycine are centrosymmetric [4-9]. In spite of this, glycine has been chosen by many research groups as a precursor material for new nonlinear material synthesis. The inappropriate choice of glycine as a precursor for NLO crystal work can be evidenced by the several improperly characterized glycine based compounds, many of which have been extensively commented in the literature [10-22]. The glycine/oxalic acid system has been the subject of recent research and a total of four compounds (Table 1), crystallizing in centrosymmetric space groups are well documented [7-9]. From this reaction system, the authors of the title paper claim to have synthesized a new NLO crystal namely gamma bis glycinium oxalate abbreviated by the code (GBGOx). In the following comment it will be shown that GBGOx is a dubious crystal and the title paper is completely erroneous.

Table 1 List of structurally characterized compounds from glycine/oxalic acid system

| Name | Formula | Space group | Ref |
|---|---|---|---|
| glycinium semioxalate (I)* | $(C_2H_6NO_2)(C_2HO_4)$ | $P2_1/c$ | 7 |
| glycinium semioxalate (II)* | $(C_2H_6NO_2)(C_2HO_4)$ | $P2_1/c$ | 9 |
| bis(glycinium) oxalate | $(C_2H_6NO_2)_2(C_2O_4)$ | $P2_1/n$ | 8 |
| bis(glycinium) oxalate methanol disolvate | $(C_2H_6NO_2)_2(C_2O_4)\cdot 2CH_3OH$ | $P2_1/c$ | 9 |

*I and II are concomitant polymorphs

**Comment**

The authors of the title paper believe that the reaction of the gamma modification of glycine (γ-glycine) with oxalic acid will result in the formation of a new NLO crystal, and hence proposed the name gamma bis glycinium oxalate (GBGOx) crystal as evidenced by their following statement, „*The crystal compound was confirmed as GBGOx. It has good agreement with BGOX parameters except space group. BGOX is crystallized as centrosymmetric space group but its analog in the case of GBGOx and it crystallized as noncentrosymmetric space group'*. Since the authors are not aware that the α- and γ-forms of glycine are commercially available, unlike the meta stable β-form, they first wanted to prepare the gamma glycine required for synthesis of this so called GBGOx. For this the authors claim to have made a homogeneous mixture containing stoichiometric ratios (1:1) of glycine and ammonium oxalate, which on slow evaporation is supposed to have yielded γ-glycine crystal. The authors claim to have analyzed by single crystal XRD and declared „*The obtained lattice parameters have confirmed glycine crystallized as gamma glycine*. Although the authors are very sure of having prepared γ-glycine, we totally disagree with their claim because the crystal isolated by the authors can not be gamma glycine but only ammonium oxalate monohydrate. In a very recent paper on the reinvestigation of a so called glycine ammonium oxalate crystal, we have shown that the slow evaporation of an aqueous solution containing equimolar quantities of γ-glycine and ammonium oxalate monohydrate results in the fractional crystallization of the less soluble ammonium oxalate monohydrate with the more soluble glycine remaining in solution [21]. The formation of ammonium oxalate monohydrate (and not γ-glycine) as the product can be readily evidenced in the reported IR spectrum in the title paper. Given the nature of expected product from the ammonium oxalate/glycine system, one finds it strange that the authors could get cell parameters which are in close agreement with that of γ-glycine. It is not clear if some convenient numbers closer to literature values were given without performing any unit cell measurement or a different crystal other than the one from the ammonium oxalate/glycine system was used for measurement. In any case the dubious nature of the single crystal work can be evidenced by the absence of a CIF file and the assignment of the chiral hexagonal space group *P3$_1$* without giving any valid reason. It is to be noted that in the original structure determination of γ-glycine, Iitaka had chosen the chiral enantiomeric space group *P3$_2$* for structure determination. It is not clear why the authors describe a *P3$_1$* space group in discussion but list a *P2$_1$/n* space group while tabulating the data.

The authors are totally unaware that the product formed in the glycine/oxalic acid system depends on the ratio of glycine:oxalic acid used for crystal growth [7-9]. In the present case the authors were also unaware that they were using ammonium oxalate monohydrate (and not glycine) for crystal growth. Hence their following claims cannot be correct. „*Again the oxalic acid is added to improve the growth condition with gamma glycine. The morphology of grown crystal changes from hexagonal to monoclinic system with space group C2 which is recognized as noncentrosymmetric so that it satisfies the basic requirements for the SHG activity of the crystal. The parameters of gamma glycine and gamma BGOx are good agreement with reported values*". The dubious nature of the single crystal X-ray result for the so called GBGOx can be evidenced by an absence of refinement details and a CIF file to support the space group assignment and the formula of *GBGOx*. The authors are unaware that in order to prove the growth of a new polymorphic modification of the well-known centrosymmetric bis(glycinium) oxalate, a rigorous structure determination is essential. Based on the reported claims in the title paper, the dubious nature of the gamma glycine crystal (which is actually ammonium oxalate monohydrate), it appears improbable that any (*C2*) unit cell was actually measured by the authors. Since ammonium oxalate monohydrate crystallizes in the orthorhombic non-centrosymmetric *P*2$_1$2$_1$2 space group [24], its presence can explain the observed SHG characteristics for the so called GBGOx crystal which is mainly ammonium oxalate monohydrate containing some impurities of oxalic acid. The spectral characterization do not merit any discussion for such dubious crystals. In summary, it can be stated that a so called gamma bis glycinium oxalate (GBGOx) crystal is an addition to the growing list of improperly characterized glycine compounds.

Footnotes:
The commercially available sample is the monohydrate